\documentstyle[sprocl,epsfig]{article}

\def\be{\begin{equation}}
\def\ee{\end{equation}}
\def\bea{\begin{eqnarray}}
\def\eea{\end{eqnarray}}

\begin{document}

\title{
\rightline{HEPSY 97-03}
\rightline{December 1997}
\quad\\
REVIEW OF EXPERIMENTAL RESULTS ON ELECTROWEAK
       PENGUIN DECAYS OF $b$ QUARK}

\author{T. SKWARNICKI}

\address{Department of Physics, Syracuse University,
\\ Syracuse, NY 13244, USA}


\maketitle\abstracts{ 
\centerline{ABSTRACT}
Status of experimental measurements of $b\to s(d)\,\gamma$, $b\to s\,l^+l^-$,
and $b\to s\,\nu\bar\nu$ is reviewed.
Future prospects are discussed.}

\vfill

To be published in the Proceedings of the Seventh
International Symposium on Heavy Flavor Physics,
University of California, Santa Barbara, California,
July 7-11, 1997.

\newpage

\section{Importance of electroweak penguin decays of $b$ quark}
\label{sec:Importance}

\begin{figure}[tbhp]
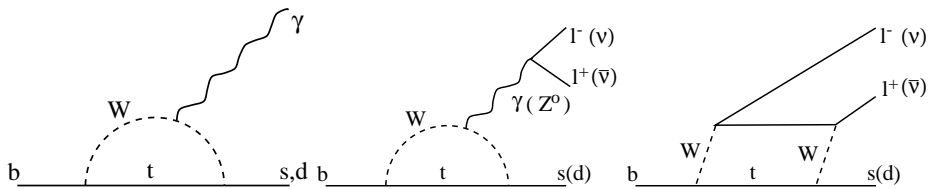

\hbox{
\psfig{figure=bsg.pstex,width=4cm}
\psfig{figure=bsll.pstex,width=4cm}
\psfig{figure=bsllbox.pstex,width=4cm}
}
\caption{Electroweak penguin and box diagrams.
\label{fig:diagrams}}
\end{figure}

Flavor Changing Neutral Currents (FCNC) are forbidden 
to first order in the Standard Model.
Second order loop diagrams (see Fig.~\ref{fig:diagrams}), known as penguin
and box diagrams, can generate 
effective FCNC which lead to $b \to s$ and $b \to d$ transitions.
Exchange of virtual top quark dominates the loop decays: $b\to t\to s (d)$.
Since the CKM matrix element $|V_{tb}|$ is very close to unity, rates
for the loop decays of $b$ quark are sensitive to $|V_{ts}|$ ($|V_{td}|$)
which will be very difficult to measure in the direct decays of the top quark.
Complementary information on $|V_{ts}|$ and $|V_{td}|$ can be
obtained from $B_s - \bar B_s$ and $B^0 - \bar B^0$ mixing.

Since the Standard Model loops involve the heaviest particles
we know to date
($t$, $W$, $Z^0$), rates for these processes are very sensitive
to possible exchange of non-standard objects like charged Higgs and other
supersymmetric particles.
Therefore, measurements of these processes constitute the most
sensitive low energy probes for certain high energy extensions
of the Standard Model.

Strange quarks can also decay through loop processes, $s \to t \to d$.
Rate for these decays is, however, strongly suppressed compared
to the loop decays of $b$ quark by unfavorable CKM elements
($|V_{ts}\cdot V_{td}|^2/|V_{tb}\cdot V_{ts}|^2\sim|V_{td}|^2\sim10^{-4}$).
Consequently, all loop decays of $s$ quark but $s\to d \,\nu\bar\nu$
are overwhelmed by long distance effects. Rare kaon experiments may
soon reach sensitivity needed to detect $s\to d \,\nu\bar\nu$ decays.

A hard 
gluon can also be emitted from the penguin loop. Even though the 
inclusive rate
for such decays is expected to be much higher than for the electroweak
penguins, many final states emerging from  $b\to s(d)\,g$ can also be
produced by tree level $b\to u$ decays. Therefore, the inclusive rate
is ill defined both theoretically (interference) and experimentally
(no common signature). Some exclusive final states uniquely identify
$b\to s\,g$ decays, but they occur at much smaller rate than
inclusive decays.
Also, the theoretical interpretation of the data is
clouded by unknown hadronization probabilities.

\section{Electromagnetic penguins}

\subsection{Exclusive $b\to s\,\gamma$ decays}
\label{sec:ksg}

Existence of the loop decays was first confirmed experimentally by 
observation of electromagnetic penguin in the exclusive mode
of $B\to K^*\,\gamma$ by CLEO-II at CESR. \cite{kstargammaPRL} 
The initial observation was based on $1.5\cdot10^6$ 
$e^+e^-\to \Upsilon(4S)\to B\bar{B}$ events.
The $K^*$ is the lightest hadron which can be produced by $b\to s\,\gamma$.
Exclusive $B$ decay 
reconstruction at the $\Upsilon(4S)$ has a very
small background thanks to the beam energy constraint: $E_B=E_{beam}$.
The detection efficiency for the 
$K^{*0}\gamma\to K^+\pi^-\gamma$ mode is 22\%\ in CLEO-II.

An updated analysis based on larger statistics 
($2.6\cdot10^6$ $B\bar{B}$ events in
$2.4$ $fb^{-1}$ of integrated luminosity)
and an improved analysis techniques 
were presented at the Warsaw conference. \cite{kstargammaWaw}
Averaged over various charge modes:
$ {\cal B}(B\to K^*\,\gamma)=(4.2\pm0.8\pm0.6)\cdot10^{-5}$.

The LEP experiments looked for these decays in $e^+e^-\to Z^0\to b\bar{b}$
but were not able to observe the signal due to an insufficient number of
$b\bar{b}$ pairs.
Hadronic colliders provide production rates superior to the ones
achievable in $e^+e^-$ collisions.
The CDF experiment at Tevatron has attempted to observe $B^0\to K^{*0}\,\gamma$
decays \cite{kstargammaCDF} 
by implementing dedicated \lq\lq penguin trigger''.
An integrated luminosity of $23.34$ $pb^{-1}$ was obtained 
yielding \footnote{To estimate number of $b\bar{b}$ pairs and 
reconstruction efficiencies for the CDF measurements, I use
$\sigma(p\bar{p}\to b\bar{b}X)=30$ $\mu b$ for $|\eta|<1$.}
about $7\cdot10^8$ $p\bar{p}\to b\bar{b}X$ events produced
in the central region ($|\eta|<1$).
The trigger required a high $P_t$ photon ($>10$ GeV) associated with
two charged tracks ($P_t>2$ GeV, $\Delta\phi<18^\circ$).
Large backgrounds from non-$b\bar{b}$ events are suppressed in
the off-line analysis by requiring a detached $B$ decay vertex and
large impact parameters  at the primary vertex of the $K^+\pi^-$ candidates.
Unfortunately the resultant experimental detection efficiency is
extremely low ($\sim0.0001\%$) 
and no signals are observed by CDF.
The upper limit set by CDF,    
$ {\cal B}(B^0\to K^{*0}\,\gamma)=<22\cdot10^{-5} $ (at 90\%\ C.L.),
is a factor of four away from the branching ratio measured by CLEO.
Using the similar analysis, CDF also sets 90\%\ C.L.\ limit:
$ {\cal B}(B_s\to \phi\,\gamma)=<39\cdot10^{-5} $,
which is only slightly looser than the limit previously obtained 
by ALEPH \cite{phigammaALEPH}: 
$<29\cdot10^{-5}$.

\subsection{Search for exclusive $b\to d\,\gamma$ decays}

Detection of  $b\to d\,\gamma$ is difficult because the rates are
suppressed by $|V_{td}|^2/|V_{ts}|^2$ $\sim10^{-2}-10^{-1}$.
Rejection of the dominant background from 
$b\to s\,\gamma$ decays requires a good
particle identification, except for the simplest exclusive final states
in which kinematic cuts alone are very effective.
CLEO-II searched for  $B\to (\rho,\omega)\,\gamma$ 
decays. \cite{kstargammaWaw}
No evidence for the signal was found due to lack of sufficient experimental
statistics ($2.6\cdot 10^6$ $B\bar{B}$ pairs).
The following upper limits were set (90\%\ C.L.):
${\cal B}(B^0\to \rho^0\,\gamma)=<3.9\cdot10^{-5} $,
${\cal B}(B^0\to \omega\,\gamma)=<1.3\cdot10^{-5} $, and
${\cal B}(B^-\to \rho^-\,\gamma)=<1.1\cdot10^{-5} $.
The ratio
${\cal B}(B\to (\rho,\omega)\,\gamma)/{\cal B}(B\to K^*\,\gamma)$
can be used to determine 
$|V_{td}|^2/|V_{ts}|^2$
after corrections for the phase space and $SU(3)-$flavor symmetry breaking
effects. Unfortunately the latter are somewhat model dependent.
Long Distance interactions may further complicate the analysis. \cite{Ali}
From the present experimental limits CLEO obtains:
$|V_{td}|^2/|V_{ts}|^2 < 0.45-0.56$,
where the range indicates the uncertainty in the theoretical factors.

With more data and improved particle identification devices
$b\to d\,\gamma$ may be observed by the next generation
of $e^+e^-\to\Upsilon(4S)$ experiments.

Another way to determine $|V_{td}|^2/|V_{ts}|^2$ is 
to use a ratio of $B^0 - \bar B^0$ and $B_s - \bar B_s$ mixing.
While $B^0 - \bar B^0$ mixing is already well measured by various
experiments, only lower bounds on $B_s - \bar B_s$ mixing have been set.
Measurement of $B_s - \bar B_s$ mixing is likely to require 
a dedicated experiment at hadronic collider with excellent time resolution
and robust triggering,
like BTeV proposed for the Tevatron, and LHC-B proposed for the LHC.

\subsection{Significance of inclusive measurements}

The measured rate for exclusive mode of $B\to K^*\,\gamma$ is in 
the ball-park of the Standard Model predictions. 
Quantitative tests of the Standard Model with rates measured
for exclusive channels are severely handicapped by our inability
to calculate hadronization probabilities from the first principles
of the theory. Predictions of phenomenological models
for $K^*$ fraction in $b\to s\,\gamma$ decays 
($R_{K^*}\equiv{\cal B}(B\to K^*\,\gamma)/{\cal B}(b\to s\,\gamma)$) 
vary in a wide range \cite{kstargammaTh}: $1-40\%$.
One should notice however, significant improvements in 
recent lattice-QCD calculations in this area. \cite{Flynn}

Fortunately, when summed over all possible final states 
hadronization probabilities drop out and inclusively measured
rate should reflect the short distance interactions
which can be accurately predicted using the effective Hamiltonian
of the Standard Model. The first non-perturbative correction 
is expected to be of second order in the
expansion over $\Lambda_{QCD}/m_b$, thus it should be small thanks
to the heavy $b$ quark mass.
Next-to-leading order perturbative calculations have been recently
completed for the $b\to s\,\gamma$. 
Assuming unitarity of the CKM matrix to constrain $|V_{ts}|$
the Standard Model predicts \cite{bsgammaTh}:
${\cal B}(b\to s\,\gamma)=(3.5\pm0.3)\cdot10^{-4}$.

\subsection{Measurement of inclusive $b\to s\,\gamma$ by CLEO}
\label{sec:cleobsg}

When reconstructing simple
exclusive final states like $B\to K^*\,\gamma, K^*\to K\pi$,
backgrounds are usually low due to the tight kinematic constraints
(here: constraints to the $B$ and $K^*$ masses, and to the beam energy).
Inclusive measurements are more challenging and they are often background
limited.

The main background limitation in CLEO comes from continuum production
of lighter quarks $e^+e^-\to q\bar q$, $q=d, u, s, c$.
These backgrounds can be reliably subtracted using data taken
below the $e^+e^-\to B\bar B$ threshold. However, statistical 
fluctuations in the background level
can easily swamp the signal unless the backgrounds are
efficiently suppressed.
Backgrounds from $B$ decays are less serious since $b\to s\,\gamma$   
decays are quasi-two-body and produce higher energy photons 
($E_\gamma\sim m_b/2$) than photons from usual decay modes.

\begin{figure}[tbhp]
\vskip-1cm
\hbox{                             \quad\hskip-3cm
\psfig{figure=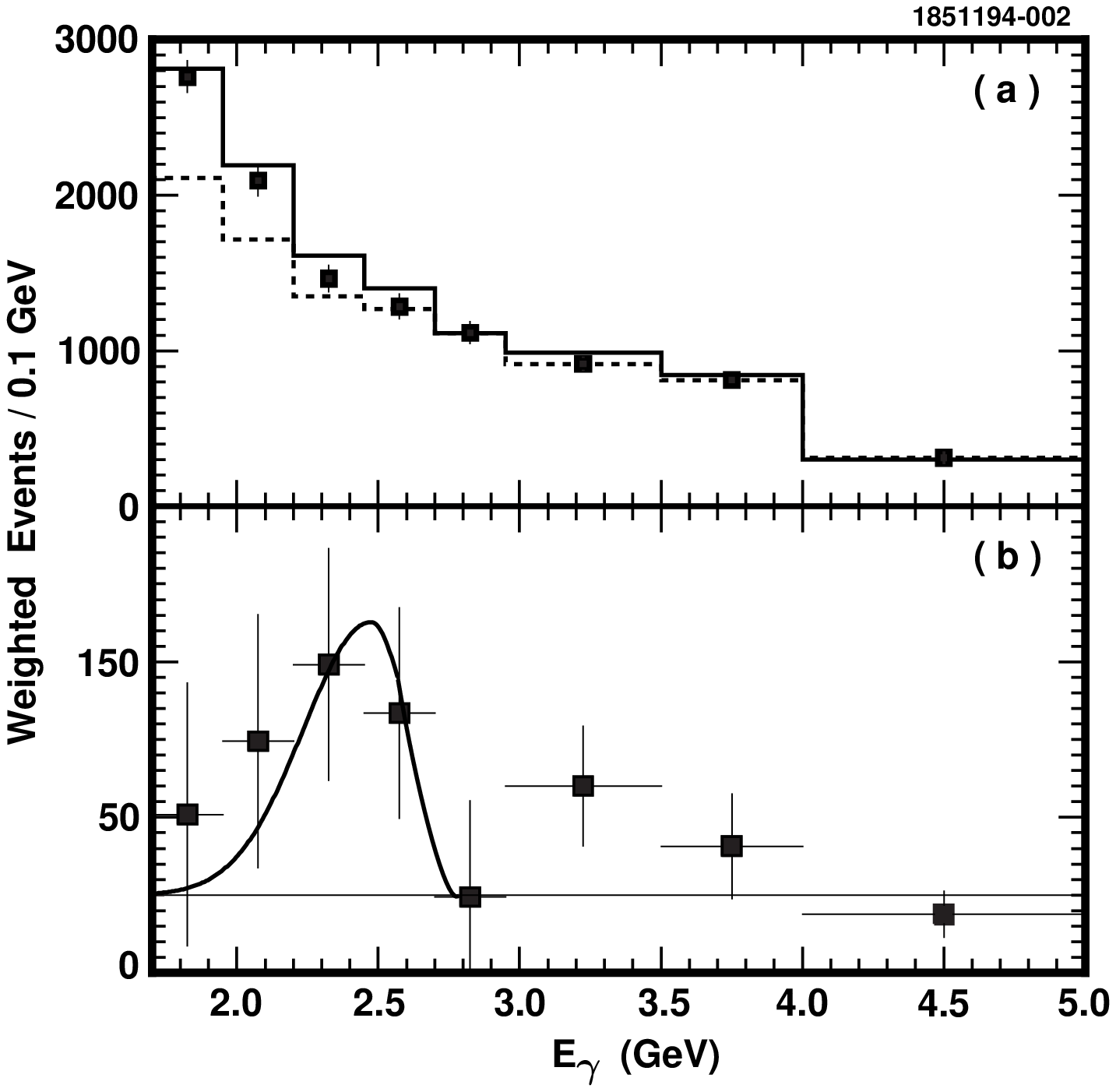,width=10cm}\hskip-3cm
\psfig{figure=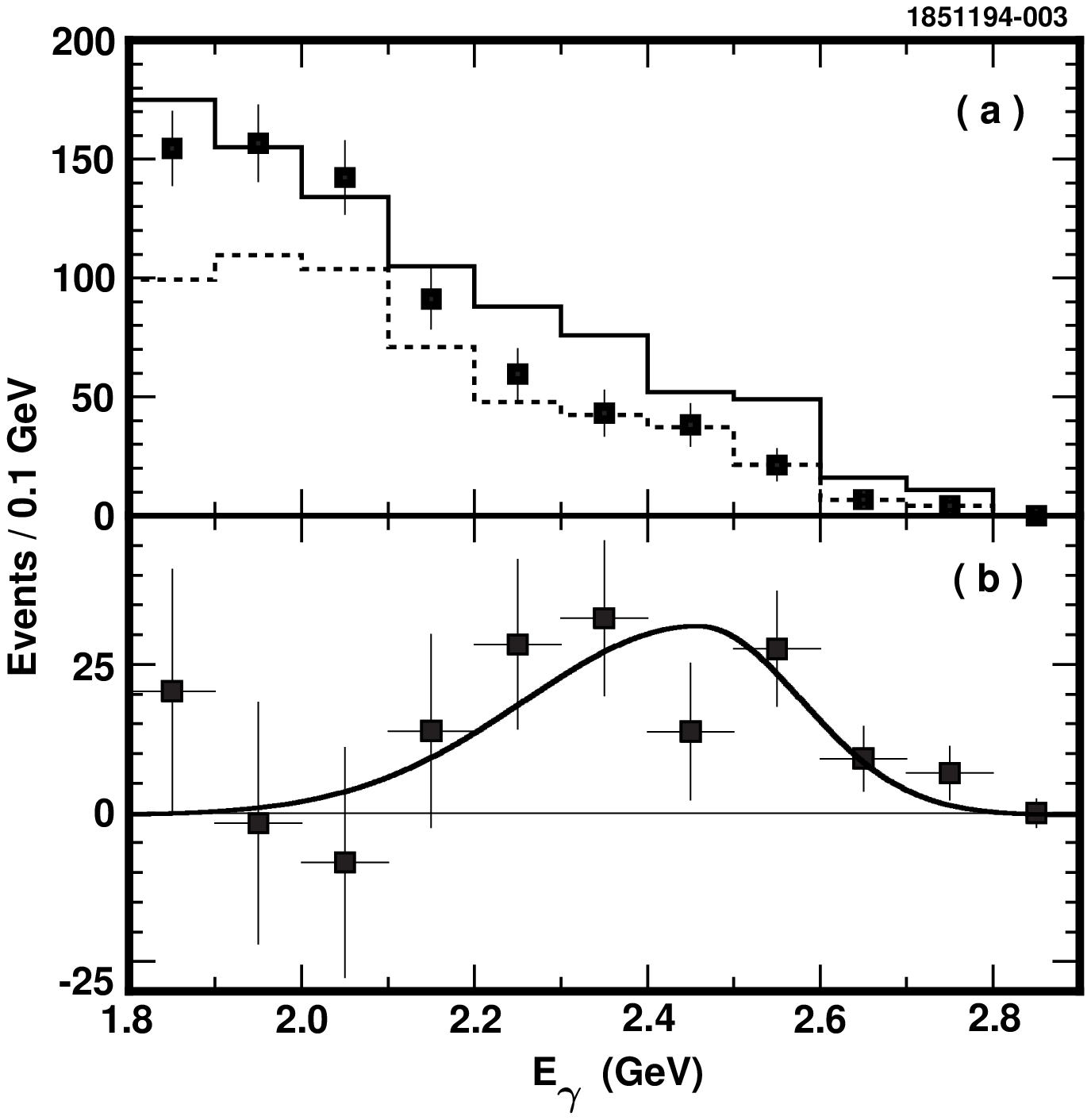,width=10cm}\hskip-3cm\quad
}
\vskip-5cm
\caption{Inclusive $E_\gamma$ spectra in the CLEO-II $b\to s\gamma$
         measurement obtained with the event-shape analysis (left)
         and with the inclusive $B-$reconstruction (right).
         (a) $\Upsilon(4s)$ data (solid histogram), scaled 
             below $\Upsilon(4S)$ data (dashed histogram)
             plus estimated $\Upsilon(4S)$ backgrounds (squares).
         (b) Background-subtracted data (points) and Monte Carlo
             prediction for the shape of the $b\to s\,\gamma$ signal
             (solid curve).
\label{fig:CLEObsg}}
%
\vskip-3cm
\hbox{
\quad\hskip-2cm
\psfig{figure=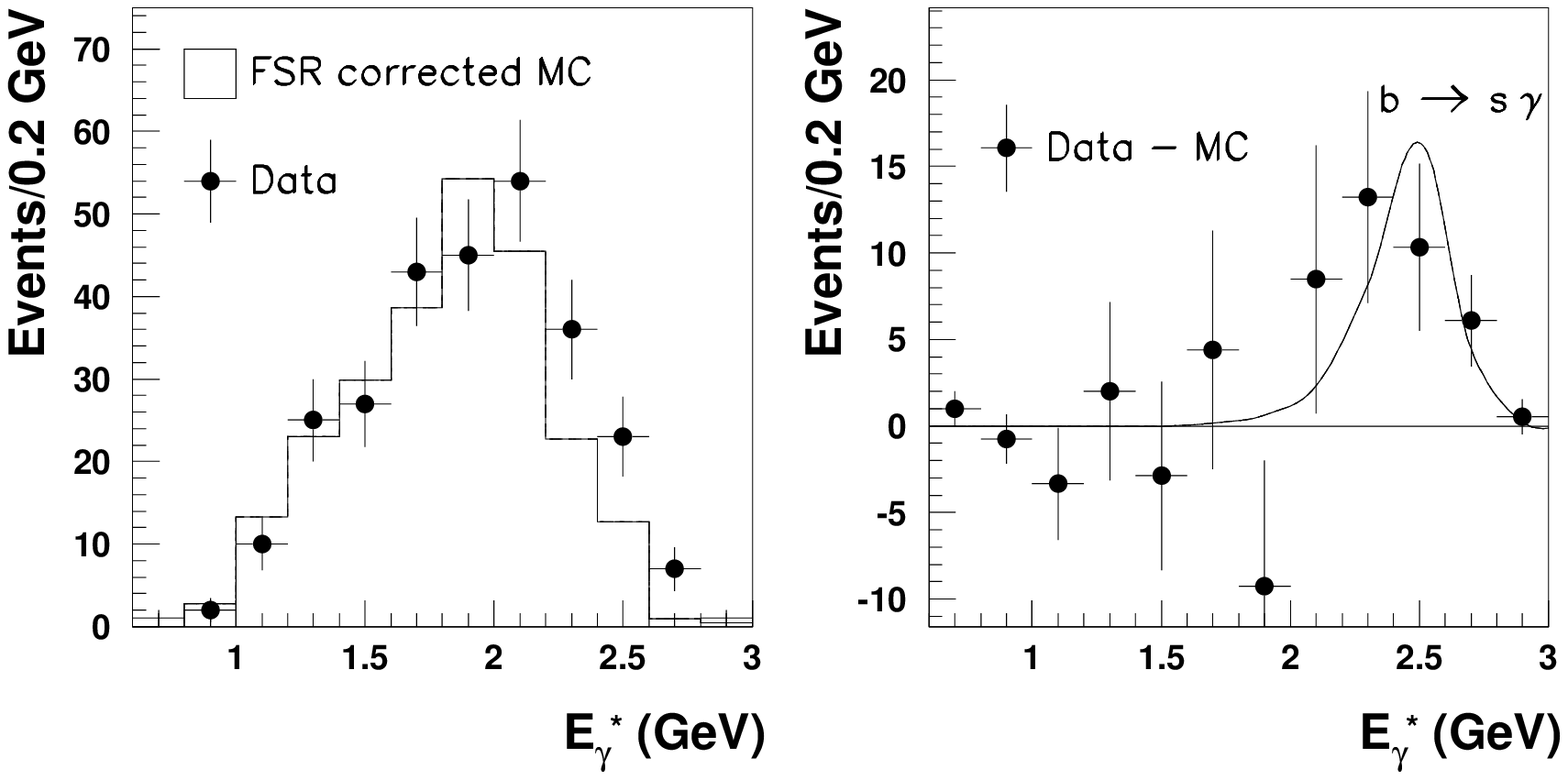,width=15cm}
\hskip-2cm\quad}
\vskip-10cm
\caption{Inclusive $E_\gamma^*$ spectrum in the ALEPH $b\to s\gamma$
         measurement. On the left: data (points) and total estimated 
         background (solid histogram). On the right: background-subtracted
         data (points) and Monte Carlo
         prediction for the shape of the $b\to s\,\gamma$ signal
         (solid curve).
\label{fig:aleph}}
\quad
\end{figure}

CLEO used two complementary
approaches to suppress the continuum background. \cite{bsgammaCLEO}
In one approach only the photon among $b\to s\,\gamma$ decay products was
explicitly reconstructed. Topological differences between $B\bar B$ events
(spherical - since $B$ mesons almost at rest at $\Upsilon(4S)$) and
$e^+e^-\to q\bar q$ events (two jets) were used for the background suppression.
For the best sensitivity all shape variables where combined using 
a neural net technique. 
The signal amplitude was extracted using a one parameter fit to the
the neural net output variable, with the signal shape and
the $B\bar{B}$ backgrounds taken from Monte Carlo
simulation, and the continuum background subtracted using the below
$\Upsilon(4S)$ data. 
In the second approach, all products of the $b\to s\,\gamma$ decay
were reconstructed as in exclusive reconstruction. Thus, the constraints
to the $B$ mass and to the beam energy could be used. 
The final state was only loosely restricted to contain a kaon 
candidate (a charged track consistent with $K^{\pm}$ by $dE/dx$ and ToF,
or a $K^0_s\to\pi^+\pi^-$ candidate) and $1-4$ pions (including at most
one $\pi^0\to\gamma\gamma$). 
The photon energy spectra measured by CLEO with these two methods in
a sample of $2.2\cdot10^6$ $B\bar{B}$ events are shown in 
Fig.~\ref{fig:CLEObsg}. 
The first method has rather large continuum background but also
high signal efficiency ($32\%$). 
The second method is very good in suppressing continuum background, but the
signal efficiency is much smaller ($9\%$).
Sensitivity of these two approaches is nearly equal, and the measurements of
signal amplitudes are only slightly correlated.
By combining these two methods together, CLEO-II
measured ${\cal B}(b\to s\,\gamma)=(2.32\pm0.57\pm0.35)\cdot10^{-4}$
in agreement with the Standard Model expectations.

Combining the inclusive and the exclusive measurements, CLEO-II determines
$R_{K^*}=(18.1\pm6.8)\%$ in agreement with 
some phenomenological estimates \cite{kstargammaTh}
and the recent QCD calculations on lattice. \cite{Flynn}

\subsection{Measurement of inclusive $b\to s\,\gamma$ by ALEPH}

Even though $e^+e^-\to b\bar{b}$ cross-section is larger at $Z^0$
than at $\Upsilon(4S)$, high luminosity is more difficult to obtain
at higher $e^+e^-$ collision energies. Thus, $b\bar{b}$ samples
obtained by the LEP experiments are smaller than the one accumulated 
at CESR. Preliminary analysis by ALEPH is based on $0.6\cdot10^6$
$b\bar{b}$ pairs. The other disadvantage for experiments at $Z^0$ 
is a loss of the beam energy constraint, since $b$ quark produces not only
a $B$ meson, but also a few fragmentation particles.
In addition to $B^-$ and $B^0$, also $B_s$ and $b-$baryons are produced.
On the other hand the $Z^0$ environment offers important advantages too.
Produced $b$ quarks are highly
relativistic, thus decay products from two $B$ mesons separate
into two back-to-back hemispheres reducing reconstruction ambiguities.
Even more important; average decay length of $B$ meson is by
two orders of magnitude larger than in CLEO (2600 $\mu m$ vs.\ 30 $\mu m$).
Therefore at the $Z^0$,
detached vertex cuts are a powerful suppression tool against
the light quark backgrounds.

ALEPH looks for inclusive $b\to s\,\gamma$ decays by combining
a high energy photon cluster with other particles in the same 
hemisphere to match the $B$ meson mass within the experimental
resolution. \cite{bsgammaALEPH} 
The energy of $B$ candidates ($E_B$) is required to be large,
since on average $B$ meson carries about 70\%\ of the beam energy.
Up to 8 particles are allowed in addition to the
photon, including charged tracks, $\pi^0$'s
detected in the electromagnetic calorimeter, and $K^0_L$ detected in
the hadronic calorimeter. To reduce confusion from the fragmentation 
pions, the charged tracks included in the $B$ meson combination must
miss the primary interaction point.
Once the $B$ candidate is created, the photon energy can be boosted
to the $B$ rest frame ($E_\gamma^*$)
where the signal photons are quasi-monochromatic: $E_\gamma^*\sim m_b/2$.
Further background suppression is achieved by requiring a detached
vertex in the opposite hemisphere. 

The selected data are fit in various bins
of $E_\gamma^*$, $E_B$, and vertex detachment in the opposite 
hemisphere allowing for three contributions: signal, final state
radiation background, and all other backgrounds. The shapes of these
contributions are fixed from the Monte Carlo simulations while
normalizations are allowed to float.
Fig.~\ref{fig:aleph} 
shows the results of the fit in a function of $E_\gamma^*$
for the tighter cuts on the other two variables. 
The total reconstruction efficiency ($\sim12\%$) is similar to the
one obtained by CLEO in their inclusive $B$ reconstruction.
In spite of four times smaller $b\bar b$ statistics, ALEPH is 
able to observe a significant inclusive signal.
This should be attributed to the better
background suppression by the vertex cuts.
The rate measured by ALEPH,
${\cal B}(b\to s\,\gamma)=(3.29\pm0.71\pm0.68)\cdot10^{-4}$,
is consistent with the CLEO measurement and the Standard Model
predictions.

The other LEP experiments were not able to see $b\to s\,\gamma$ signal
and set upper limits consistent with the CLEO and ALEPH measurements:
\newline
DELPHI \cite{bsgammaDELPHI}  $<5.4\cdot10^{-4}$, 
L3 \cite{bsgammaL3} $<12\cdot10^{-4}$ (90\%\ C.L.)

\subsection{Theoretical implications of the inclusive measurements}

Combining the CLEO measurement with the preliminary ALEPH result
and dividing by the Standard Model predictions Ali \cite{Ali}
obtains:
$$\left|\frac{V_{ts}^*}{V_{cb}} V_{tb}\right| 
=0.84\pm0.09\hbox{(experiment)}\pm0.04\hbox{(theory)}$$
consistent with the unitarity constraint \cite{Ali}:
$$\left|\frac{V_{ts}^*}{V_{cb}} V_{tb}\right|\approx|V_{cs}|=1.01\pm0.18$$
Using the measured values to eliminate 
$V_{tb}$ and $V_{cb}$, Ali extracts \cite{Ali}
from the $b\to s\,\gamma$ measurements:
$$|V_{ts}|=0.033\pm0.007$$

The agreement between the measured and the Standard Model rates
(including the CKM matrix unitarity) 
leaves a little room for non-standard contributions.
Meaningful constraints on many extensions of the Standard Model
can be obtained as discussed by J. Hewett at this
conference. \cite{Hewett}

\subsection{Future prospects}

The theoretical uncertainties in the predictions for
the inclusive $b\to s\,\gamma$ rate are smaller than 
the present experimental errors, calling for improved
measurements.
The CLEO experiment has more than doubled their data sample
since the first measurement of the inclusive $b\to s\,\gamma$ rate,
and it still accumulates the data.
The data analysis is under progress.
No more data is expected at the $Z^0$ peak at LEP.
In the future, $\Upsilon(4S)$ $B-$factories 
at CESR (CLEO-III experiment), PEP-II (BaBar experiment) 
and KEK-B (Belle experiment)
will produce very large samples of $b\to s\,\gamma$ events
with smaller backgrounds thanks to the improved particle identification
(all three experiments) and some $B$ vertex capability (PEP-II and KEK-B).
Photon energy spectrum should be measured with a good accuracy.
Detection of $b\to d\,\gamma$ via inclusive $B$ reconstruction may not
be out of question.
Potential of future hadronic collider experiments for electromagnetic
penguins has not been explored.
The CDF has made the first step in this direction as discussed in
section. \ref{sec:ksg}
If photons detected in an 
electromagnetic calorimeter yield too much background, use of
converted photons which can be pointed to a detached $B$ vertex should
be investigated.

\section{Searches for $b\to s\,l^+l^-$}

\subsection{Theoretical expectations}

The  $b \to s l^+ l^-$  decay rate 
is expected in the Standard Model
to be nearly two orders of magnitude lower than the rate for
$b \to s \,\gamma$ decays. \cite{Ali,refAH}
Nevertheless, 
the $b \to s l^+ l^-$ process has received considerable
attention since it offers a deeper insight into 
the effective hamiltonian
describing FCNC processes in $B$ decays. \cite{Ali}
While $b \to s \,\gamma$ is only sensitive to the
absolute value of the $C_7$  
Wilson coefficient 
in the effective hamiltonian,
$b \to s l^+ l^-$ is also sensitive
to the sign of $C_7$ and
to the $C_9$ and $C_{10}$ coefficients, 
where the relative contributions vary with $l^+l^-$ mass.
These three coefficients are related to three 
different processes contributing to $b \to s\, l^+l^-$: 
$b\to s(\gamma\to l^+l^-)$, $b\to s (Z^0\to l^+l^-)$, 
and box diagram (see Fig.~\ref{fig:diagrams}).
Processes beyond the Standard Model 
can alter both the magnitude and the sign
of the Wilson coefficients.

\subsection{Searches in exclusive modes}

The simplest allowed final states are 
$B\to K \,l^+l^-$, and $B\to K^* \,l^+l^-$ 
($B\to K\,\gamma$ is forbidden by the angular momentum
conservation). Each of them is expected to constitute
$\sim10\%$  of the total $b\to s \,l^+l^-$ rate.
The most sensitive searches for these decays were 
performed by CDF and CLEO-II experiments.

\begin{figure}[p]
\vskip-7cm
\hbox{                             \quad\hskip-2cm
\psfig{figure=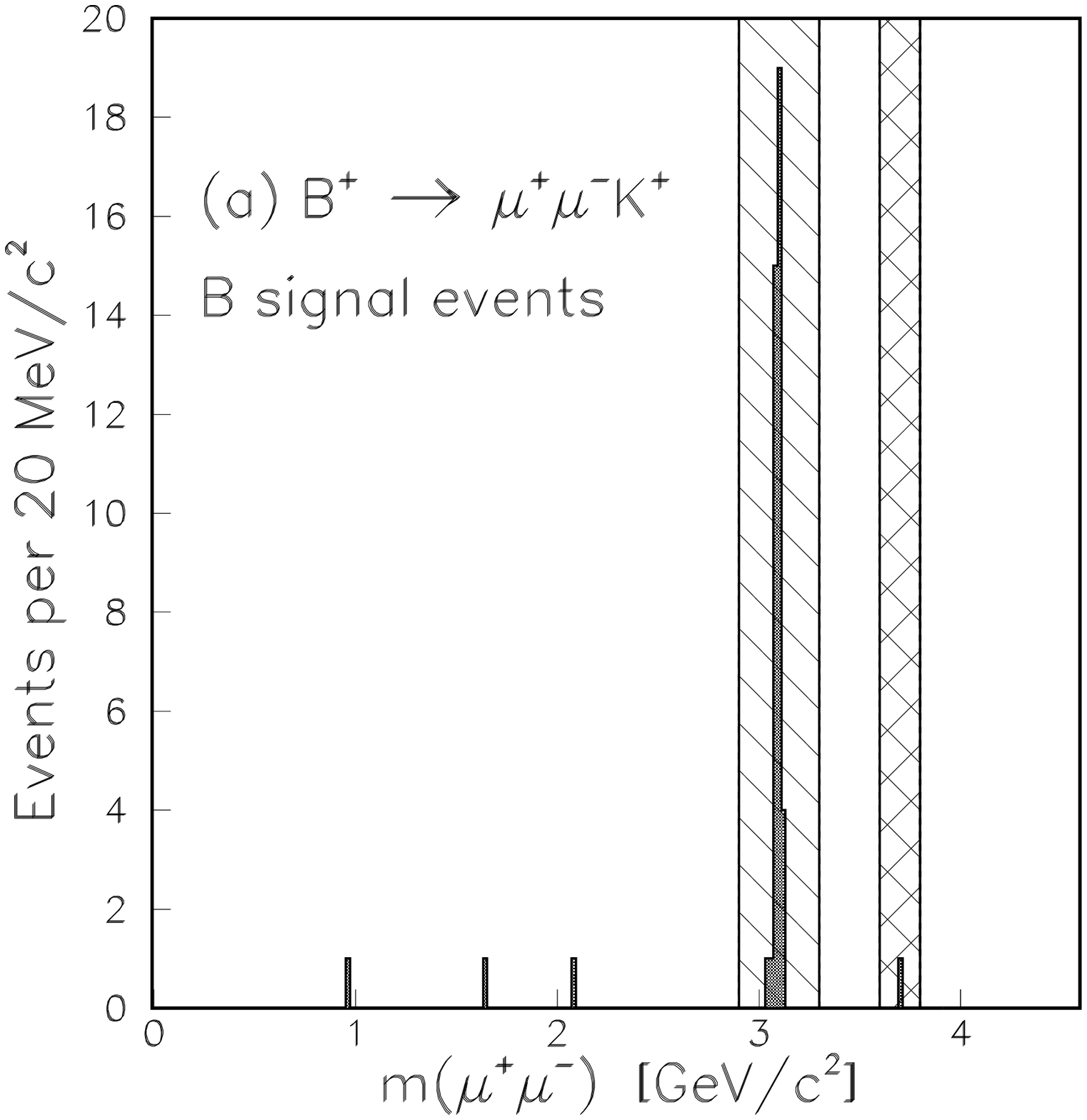,width=9cm}\hskip-1cm
\psfig{figure=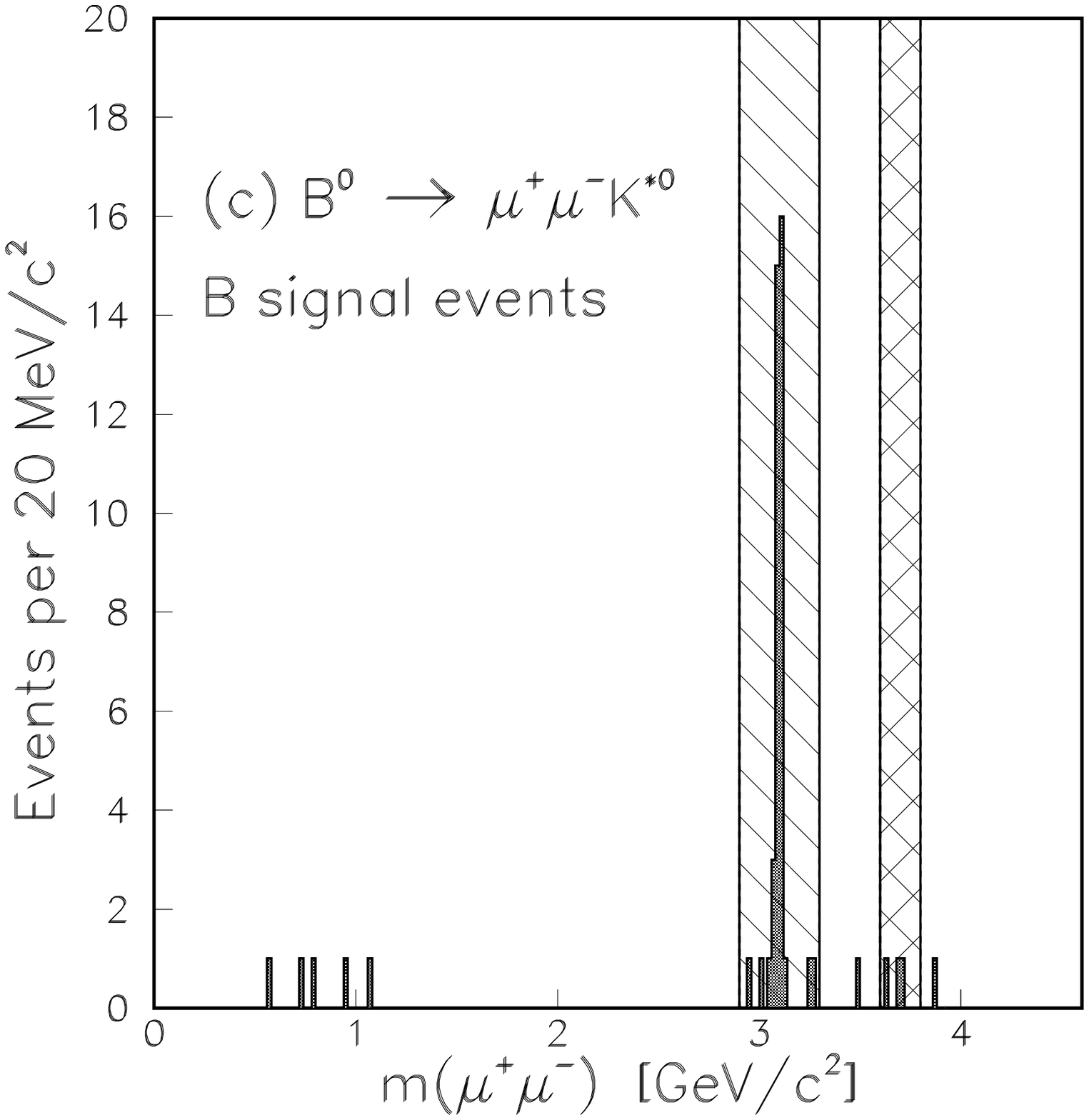,width=9cm}\hskip-2cm\quad
}
\caption{Di-muon mass distributions in the CDF 
         search for $b\to s\,\mu^+\mu^-$ via {\bf exclusive}
         final states $B^+\to K^+\,\mu^+\mu^-$ (left) and
         $B^0\to K^{*0}\,\mu^+\mu^-$ (right).
\label{fig:CDFsll}}
%
\hbox{
\quad\hskip-2cm
\psfig{figure=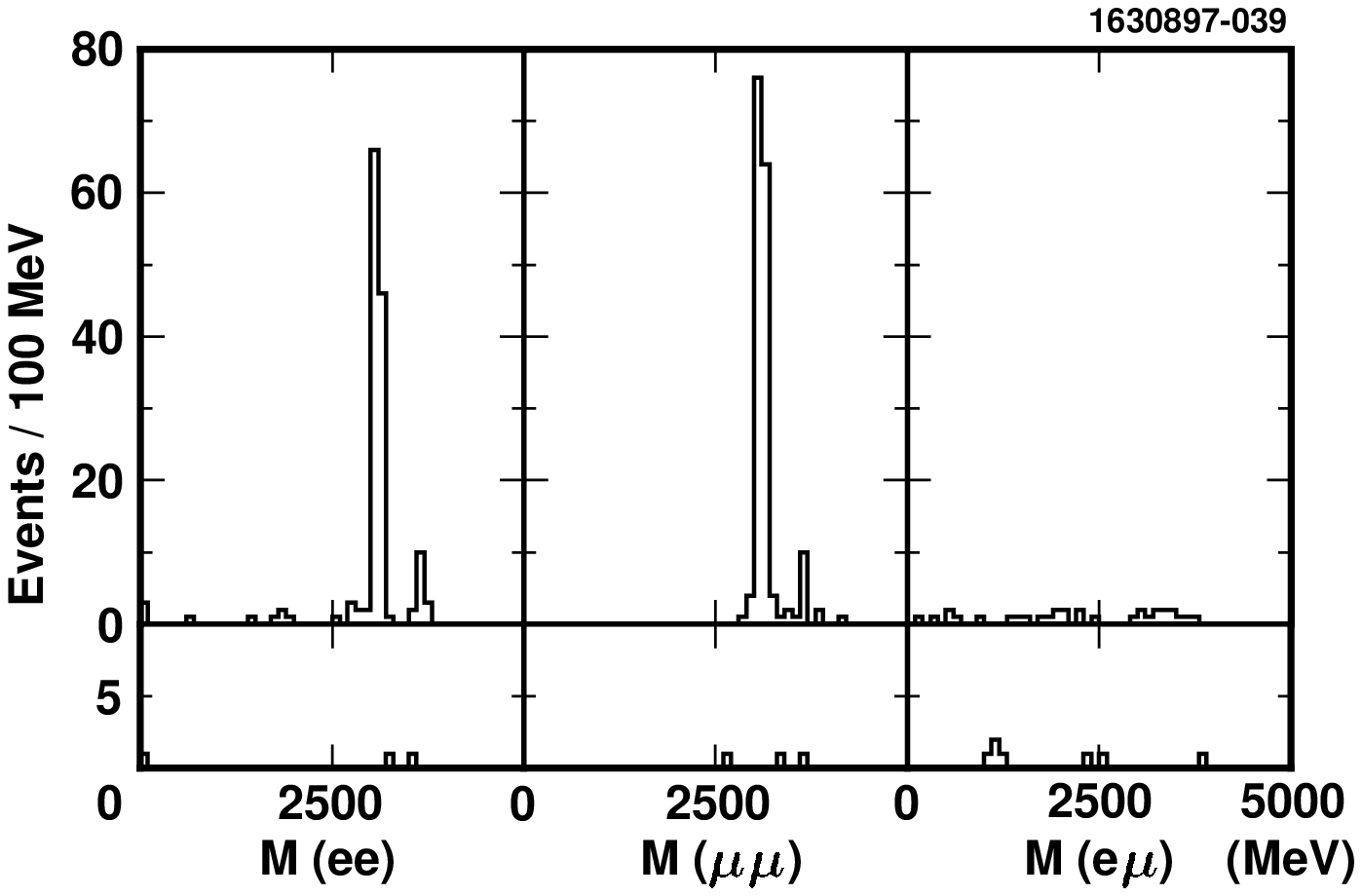,width=16cm,height=12cm}
\hskip-2cm\quad}
\vskip-3cm
\caption{Di-lepton mass distribution in the CLEO-II {\bf inclusive}
         search for $b\to s\,l^+l^-$. 
        On- (top) and off- (bottom) $\Upsilon(4S)$ data are shown.
\label{fig:CLEOsll}}
\quad
\end{figure}

The CDF search \cite{sllCDF} 
is based on 17.8 pb$^{-1}$ of data 
($\sim5\cdot10^8$ $b\bar b$ pairs for $|\eta|<1$)
and di-muon trigger. The backgrounds are suppressed
by transverse momentum cuts ($P_t(\mu)>2, 2.5$ GeV,
$P_t(K^{(*)})>2$ GeV, $P_t(B)>6$ GeV), detached vertex
cut ($c\tau(B)>100 \mu m$), isolation requirement and
$B$ mass cut. The resulting di-muon mass distributions
are shown in Fig.~\ref{fig:CDFsll}.
The signals due to long distance interactions
$B\to K^{(*)}\psi^{(')}$ are observed. 
Since the branching ratios for these decays were previously 
measured by the other experiments, CDF used these
signals for normalization.
Reconstruction efficiencies are roughly
$0.13\%$ for the $K$, and $0.07\%$ for the $K^*$ modes.
A few events observed outside the $\psi$ and
$\psi'$ bands are consistent with the background
estimates. The following upper limits are set:
${\cal B}(B^-\to K^-\,\mu^+\mu^-)<1.0\cdot10^{-5}$ and
${\cal B}(B^0\to K^{*0}\,\mu^+\mu^-)<2.5\cdot10^{-5}$ 
(90 \%\ C.L.).

The CLEO-II experiment searched for these decays
in a sample of $b\bar b$ pairs 
by two orders of magnitude smaller 
($\sim2.2\cdot10^6$ $B\bar{B}$) than in the CDF analysis,
though with efficiencies larger also by two orders of 
magnitude ($\sim15\%$ for $K$ and $\sim5\%$ for $K^*$),
and suitably low backgrounds.
Thus, by this strange coincidence the sensitivity of the
CDF and of the CLEO-II experiments were very similar.
In addition to the limits in the di-muon mode,
${\cal B}(B^-\to K^-\,\mu^+\mu^-)<0.9\cdot10^{-5}$ and
${\cal B}(B^0\to K^{*0}\,\mu^+\mu^-)<3.1\cdot10^{-5}$,
CLEO also set the limits using di-electrons:
${\cal B}(B^-\to K^- \,e^+e^-)<1.2\cdot10^{-5}$ and
${\cal B}(B^0\to K^{*0} \,e^+e^-)<1.6\cdot10^{-5}$,
 
The experimental limits are an order of magnitude 
away from the Standard Model predictions.

\subsection{Inclusive searches}

The new CLEO analysis \cite{sllCLEO} 
looks for inclusive $b\to s \,l^+l^-$ decays using the inclusive $B$
reconstruction technique previously described
for the $b\to s\,\gamma$ decays (see section~\ref{sec:cleobsg}).
The obtained di-lepton mass spectra
are shown in Fig.~\ref{fig:CLEOsll}.
Again clear signals for $B\to X_s\psi$ and
$B\to X_s\psi'$ are observed. Events outside the
$\psi$ and $\psi'$ bands are consistent with the
$B\bar{B}$ background estimates (the continuum backgrounds
are small). With a sample of $3.3\cdot10^6$ $B\bar{B}$
pairs and reconstruction efficiencies around $5\%$,
CLEO sets the 90\%\ C.L.\ upper limits,
${\cal B}(b\to s \,e^+e^-)<5.7\cdot10^{-5}$ and
${\cal B}(b\to s \,\mu^+\mu^-)<5.8\cdot10^{-5}$
(combined: ${\cal B}(b\to s \,l^+l^-)<4.2\cdot10^{-5}$),
which are again an order of magnitude away from the
Standard Model predictions \cite{refAH}:
${\cal B}(b\to s \,e^+e^-)=(0.8\pm0.2)\cdot10^{-5}$ and
${\cal B}(b\to s \,\mu^+\mu^-)=(0.6\pm0.1)\cdot10^{-5}$.

The upper limits on inclusive $b\to s\,\mu^+\mu^-$
previously presented by the experiments 
at hadronic colliders \cite{UA1,D0}
appear to be based on overestimated sensitivity.
\footnote{
The results were obatined by UA1 \cite{UA1}
at $Sp\bar{p}S$ and D0 at Tevatron. \cite{D0}
I have simulated efficiency of the UA1 kinematic cuts with
PYTHIA and the modern $b\to s\,\mu^+\mu^-$ theory and 
obtained a number by a factor of three lower than 
the overall efficiency estimated by UA1.
Allowing for the trigger and reconstruction losses
makes the discrepancy larger. 
I find the $B\bar{B}$ background subtraction method used by UA1
highly questionable as well.
The preliminary result reported
by D0 at the Warsaw conference~\cite{D0}
is currently under revision.~\cite{D0priv}}
Therefore, they are not included here.

\subsection{Future prospects}

The search for exclusive channels by CLEO-II was 
statistics limited. 
An order of magnitude increase in $B\bar{B}$ statistics
expected for the CLEO-III phase of the CESR program
should put these channels in detectable range.
However, measurement of the inclusive rate is
questionable since the CLEO-II results are already background
limited due to random combinations of leptons from
semileptonic $B$/$D$ decays and of the other particles
from two $B$ mesons in the event.
Asymmetric $B$ factories at SLAC and KEK may be better
suited for suppression of these backgrounds.
The interesting physics lies not only in the measurement
of total rates, but also in studies of di-lepton mass
distribution and forward-backward lepton charge asymmetry.
Such studies will require huge experimental
statistics which will be very difficult to achieve
at $e^+e^-$ colliders. Very large statistics will also
be required for detection of any $b\to d \,l^+l^-$ decays.
Thus, more detailed exploration of these decays will be
performed at hadronic colliders.
Triggering on di-leptons is relatively easy.
The upgraded CDF and D0 experiments should be able
to observe the exclusive modes during the Main 
Injector Run at Tevatron.
Measurement of the inclusive rate and studies of 
the differential distributions for di-leptons 
will likely require good $K/\pi$ separation
and excellent vertex resolution. 
These are attributes of the
BTeV and LHC-B experiments.

\section{Searches for $b\to s\,\nu\bar\nu$}

\begin{figure}[tbhp]
\quad\hskip3cm\psfig{figure=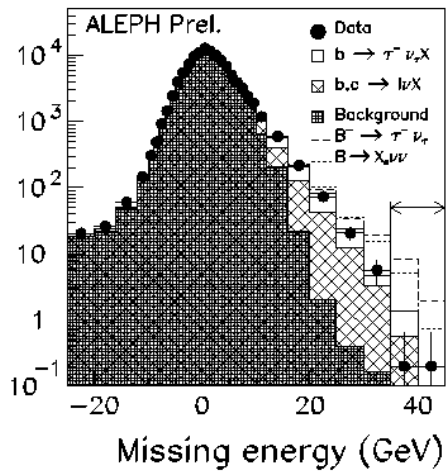,width=6cm}
\caption{Missing energy in a hemisphere for the selected $b\bar{b}$
         events by ALEPH (points). Shaded histograms show 
         the estimated background distribution. The expected 
         $b\to s\,\nu\bar\nu$ signal shape is indicated by 
         dotted line. The two highest bins are used to set
         the upper limit.
\label{fig:nunu}}
\end{figure}

The rate for  $b\to s\nu\bar \nu$
is enhanced compared to the
$b\to s \,l^+l^-$ decays primarily
by summing over three neutrino flavors 
($b\to s\tau^+\tau^-$ has a small expected rate and
will be difficult to detect experimentally).
The predicted rate is only a factor of ten lower than
for $b\to s\,\gamma$ \cite{BurasWaw}: $(3.8 \pm 0.8)\cdot 10^{-5}$.  
In principle, these decays are the cleanest theoretically 
among all penguin decays.
Therefore, measurement of inclusive rate for this
process would be of considerable interest.
Unfortunately, the neutrinos escape the detection
making it difficult for experimentalists to control
the backgrounds. 
So far, only LEP experiments have been able to
probe these decays by requiring a very large 
missing energy in a hemisphere. \cite{nunuALEPH,nunuThE} 
Semileptonic backgrounds are reduced by eliminating events
with identified lepton in the signal hemisphere.
Detached vertex in the opposite hemisphere 
suppresses non-$b\bar b$ backgrounds.
Missing energy distribution in a $b-$hemisphere
obtained by ALEPH \cite{nunuALEPH}
in a sample of $\sim0.5$ $b\bar{b}$ pairs
is shown in Fig.~\ref{fig:nunu}.
From the lack of excess of events over the semileptonic
backgrounds at the highest energy bins, ALEPH obtained:
${\cal B}(b\to s\,\nu\bar\nu)<7.7\cdot10^{-4}$ at 90\%\ C.L.

Exclusive mode of $B\to K^*\,\nu\bar\nu$ should 
constitute about 30\%\ of the total rate. \cite{am}
DELPHI has set the following upper limits \cite{bsgammaDELPHI}:
${\cal B}(B^0\to K^{*0}\,\nu\bar\nu)<1.0\cdot10^{-3}$ 
and
${\cal B}(B_s\to \phi\,\nu\bar\nu)<5.4\cdot10^{-3}$
(90\%\ C.L.)

The inclusive limit set by ALEPH is an order of magnitude
away from the expected rate.
Unfortunately, no more data is expected at the $Z^0$ peak
at LEP. 
Perhaps, $\Upsilon(4S)$ experiments will be able to
develop analysis techniques which will probe these decays
in the future high statistics data.
It is hard to imagine that experiments at hadronic
colliders will ever have any sensitivity to these decays.

\section{Summary}

Among all electroweak penguin processes only
$b\to s\,\gamma$ has been detected experimentally.
CLEO-II measured the inclusive rate and the exclusive
fraction for $B\to K^*\,\gamma$, both with about 30\%\
accuracy.
Recently, the ALEPH experiment has also observed the inclusive
signal with the rate consistent with CLEO.
Incremental increase of statistics by CLEO-II experiment
will soon allow reduction of experimental errors by up to
a factor of 2. In a few years,
further improvements are expected by CLEO-III, BaBar and
Belle experiments.

Experimental upper limits on all other penguin processes
are roughly an order of magnitude away from the Standard
Model predictions.

Exclusive $b\to d\,\gamma$ and $b\to s \,l^+l^-$ decays
should be detectable by the next generation of $e^+e^-$
experiments. The latter should also be observable
by the central detectors during the Run II of the 
Tevatron collider. 
Detailed exploration of $b\to s \,l^+l^-$ processes
(inclusive rate, di-lepton mass distribution, 
$b\to d \,l^+l^-$) are likely to require 
specialized experiments at a hadronic collider
with good kaon identification and excellent vertex
resolution, like in the proposed BTeV and LHC-B detectors.
   
The $b\to s\,\nu\bar\nu$ decays are the cleanest theoretically
but the hardest experimentally. They may remain
undetected for a foreseeable future.

\section*{Acknowledgments}

I would like to thank Ahmed Ali and
Andrzej Buras for helpful discussions of theoretical
issues. 
I would also like to thank experimentalists from
various collaborations for help in collecting
the results (Roger Forty, Joseph Kroll, Karen Lingel, 
Joachim Mnich, Franco Simonetto, and Andrzej Zieminski).

\section*{References}

\def\etal{{\it et al.}}

\end{document}